# Isotropic or anisotropic screening in black phosphorous: can doping tip the balance?

Zhimin Liu[1], Ye Yang[1], Yueshao Zheng[1], Qinjun Chen[1], Yexin Feng[1, †]

[1] *Hunan Provincial Key Laboratory of Low-Dimensional Structural Physics & Devices, School of Physics and Electronics, Hunan University, Changsha 410082, People's Republic of China*

*Corresponding author. E-mail:* [†]*yexinfeng@pku.edu.cn*

Black phosphorus (BP), a layered van der Waals (vdW) crystal, has unique in-plane band anisotropy and many resulting anisotropy properties such as the effective mass, electron mobility, optical absorption, thermal conductivity and plasmonic dispersion. However, whether anisotropic or isotropic charge screening exist in BP remains a controversial issue. Based on first-principles calculations, we study the screening properties in both of single-layer and bulk BP, especially concerning the role of doping. Without charge doping, the single-layer and bulk-phase BP show slight anisotropic screening. Electron and hole doping can increase the charge screening of BP and significantly change the relative static dielectric tensor elements along two different in-plane directions. We further study the charge density change induced by potassium (K) adatom near the BP surface, under different levels of charge doping. The calculated two-dimensional (2D) charge redistribution patterns also confirm that doping can greatly affect the screening feature and tip the balance between isotropic and anisotropic screening. We corroborate that screening in BP exhibit slight intrinsic anisotropy and doping has significant influence on its screening property.

# 1 Introduction

BP, as a typical layered vdW crystal, has attracted great attentions due to its high carrier mobility, widely tunable bandgap and remarkable in-plane anisotropic properties [1-12]. Each phosphorus atom of BP covalently bonds to three neighboring atoms, and forms non-equal bond lengths and bond angles along its orthogonal directions [13,14]. The unique low-symmetry puckered structure leads to substantial in-plane anisotropy such as the effective mass, electron mobility, optical absorption, thermal conductivity and plasmonic dispersion [15-19]. The anisotropy properties of BP have a wide range of potential applications, from BP-based plasmonic devices to high performance thin-film electronics, as well as mid- and near-infrared optoelectronics and so on [2,20-22].

Although almost all reported physical properties of BP show anisotropy in the basal plane, whether anisotropic or isotropic charge screening exist in BP is still under debate [16]. Based on effective low-energy Hamiltonian model, Tony Low and coworkers found that the in-plane static screening in BP remains relatively isotropy [17]. Beyond the long-wavelength limit, D. A. Prishchenko *et al.* demonstrated that dielectric function of few-layer BP exhibits strongly anisotropic behavior with tight-binding model and rigorously determined bare Coulomb interactions [23]. Experimental studies also report contrasting results on this issue. Tian and coworkers used the tip-induced band bending of a scanning tunneling microscope (STM) to measure the Coulomb field of ionized K adatoms on BP [16]. They observed an isotropic electrostatic screening of point charges at BP surface [16]. However, another experiment measured the ordering patterns of adsorbed K atoms on BP surface with STM and reported that the screening property of BP shows strong anisotropy [24]. They also found that, with the increase of adatom density at the BP surface, the changes of K ordering patterns suggest a transition from isotropic to anisotropic screening in BP [24]. Besides, there are also studies showing that doping has an impact on the dielectric functions and planar average of the charge density in some other kinds of semiconductors [17,25-26].

In this work, we study the charge screening in BP and its relationship with charge doping by performing first-principles calculations. We calculated the dielectric tensor elements parallel to *a* ($\varepsilon_a$), *b* ($\varepsilon_b$), and *c* ($\varepsilon_c$) directions. The calculated static dielectric tensor elements show little difference along two in-plane directions, where $\varepsilon_b/\varepsilon_a$ is 1.14 for bulk BP and 1.03 for single-layer BP, suggesting slight anisotropic screening behavior. In the following discussion of this letter, "slight anisotropy" applies to the situation when the ratio of $\varepsilon_b$ to $\varepsilon_a$ is in the range of 0.75~1.25 for bulk BP and 0.9~1.1 for single-layer BP. Frequency-dependent dielectric functions in the long-wavelength limit also show nearly isotropic characteristic. Increasing momentum can lead to anisotropic dielectric curves along with the frequency, which agrees with previous results that the dispersion of the plasmons versus momentum for BP shows strong anisotropic property. We then study the charge screening in BP under different levels of charge doping. The calculated static dielectric tensor elements with different degrees of doping indicate that electron and hole doping obviously increase charge screening and change the relative static dielectric tensor elements between two in-plane directions. The simulated charge density redistribution patterns induced by K adatom on the BP surface also confirm that the charge doping would have a significant impact on the screening in BP.

## 2 Calculation methods

Our structure optimizations and band structure calculations were performed with density function theory (DFT) as implemented in Quantum Espresso (QE) [27]. The exchange–correlation functionals were described by Perdew-Burke-Ernzerhof (PBE) functional within the framework of generalized-gradient approximation (GGA) [28]. The experimental lattice parameters of bulk BP [29] were used for band structure and screening property simulations. We built the single-layer BP model by cleaving bulk BP structure along the [001] orientation, with a vacuum space of 15.5 Å in the z-direction. The in-plane lattice parameters are obtained with PBE functional. The electronic wavefunctions were expanded with kinetic energy cut-off of 20.6 Ry and the k-mesh of 10×12×4 was used. Convergence threshold for total energy was set to be $10^{-9}$ Ry and the threshold of force minimization was $5\times10^{-4}$ Ry/Bohr.

Static dielectric tensor elements for BP as a function of doping concentration was calculated within Berkeley GW [30]. Through computing electronic eigenvalues and eigenvectors, the static or dynamic polarizability and corresponding inverse dielectric function within the random-phase approximation (RPA) could be calculated [30]. The static RPA polarizability is obtained by

$$\chi_{GG'}(\mathbf{q};0) = \sum_n^{occ} \sum_{n'}^{emp} \sum_{\mathbf{k}} M_{nn'}^*(\mathbf{k},\mathbf{q},\mathbf{G}) M_{nn'}(\mathbf{k},\mathbf{q},\mathbf{G}') \frac{1}{E_{n\mathbf{k}+\mathbf{q}} - E_{n'\mathbf{k}}} \quad (1),$$

with the plane-wave matrix elements written as

$$M_{nn'}(\mathbf{k},\mathbf{q},\mathbf{G}) = \langle n\mathbf{k}+\mathbf{q}|e^{i(\mathbf{q}+\mathbf{G})\cdot \mathbf{r}}|n'\mathbf{k}\rangle \quad (2),$$

where $\mathbf{q}$ represents a vector in the first Brillouin zone, and $\mathbf{G}$ means a reciprocal-lattice vector. $\langle n\mathbf{k}|$ and $E_{n\mathbf{k}}$ are the mean-field electronic eigenvectors and eigenvalues. Then, it is straightforward to determine the RPA dielectric matrix as

$$\epsilon_{GG'}(q;0) = \delta_{GG'} - v(\mathbf{q}+\mathbf{G})\chi_{GG'}(\mathbf{q};0) \quad (3).$$

Considering the metallicity induced by charge doping, we treated the doped BP as metal system and used intensive k-point meshes to calculate the static dielectric tensors.

Charge density difference induced by K adatom at the single-layer BP surface was simulated by using Vienna *ab initio* simulation package (VASP) [31-32]. During the calculations, we used a supercell with 10×8 surface periodicity and a vacuum space of 15.5 Å.

## 3 Results and discussion

In this work, we take single-layer BP and bulk BP as the prototypes to study the charge screening property of BP. In Fig.1, we present their crystal structures, Brillouin zones and electronic band structures. BP has a puckered orthorhombic structure which belongs to the $D_{2h}$ point group [2]. As shown in Fig. 1(d), each phosphorus atom is bound to three atoms, forms a ridge structure along the zigzag (*a*) direction and a puckered structure along the armchair (*b*) direction [13,14]. BP's structure exhibits low-symmetry due to the non-equal bond lengths and bond angles along its orthogonal directions [14,15]. The corresponding band structures of BP are shown in Fig. 1(a)-(c), in which PBE method and experimental lattice constants are used [29]. BP was reported to have a direct band gap of 0.31~0.35 eV which changes with its thicknesses [33]. One can see that valence band minimum (VBM) and conduction band maximum (CBM) appear at the same high-symmetry point in each calculated band structure, which indicates the direct band gap of BP. However, there is an overlap between VB and CB in the band structure of bulk BP shown in Fig. 1(a), suggesting that bulk BP possesses artificial metallic property. This phenomenon is induced by the underestimation of band gap by the PBE method. This artificial metallic feature will lead to serious error in evaluating dielectric constants of BP as discussed below. Therefore, we have calculated the dielectric tensor elements along *a*, *b* and *c* directions and the band structures of bulk BP with different strains. Band gap opening could be observed when in-plane strain increases to 2% and the band gap with 3% in-plane strain is close to the experimental one [33]. As shown in Table S1, $\varepsilon_b$ is very large with the in-plane strain less than 2%. While the in-plane strain become larger than 3%, the dielectric tensor elements along three directions are close to each other. So, we applied 3% strain to the *a* and *b* directions of bulk BP to obtain more realistic electronic structures. As shown in Fig. 1(b), the band gap opening could be observed with a band gap of ~ 0.1 eV. In following calculations and discussion of screening in bulk BP, we use this 3% in-plane strain model to avoid the artificial metallic behavior. In Fig. 1(c), the band gap for single-layer BP is ~ 0.9 eV, which is a bit smaller than the GW calculation result [33], but will not affect the calculated dielectric behavior much.

Static dielectric tensor elements parallel to *a* ($\varepsilon_a$), *b* ($\varepsilon_b$), and *c* ($\varepsilon_c$) directions under long-wavelength limit were calculated by PBE method. The results are listed in Table I. Ideal 2D materials (0 thickness) should possess static dielectric constants of 1. Here the calculated dielectric tensor elements in *a* and *b* directions of single-layer BP are both close to 1. Moreover, single-layer BP exhibits slight anisotropic charge screening along two directions. Bulk BP without strain shows obvious anisotropic screening. But the very large $\varepsilon_b$ comes from the artificial band inversion along G-Y direction in Brillouin zone. While in bulk BP with 3% in-plane strain, the difference of $\varepsilon_a$ and $\varepsilon_b$ is obviously smaller, which is in agreement with the experimental results reported in Ref. [34]. So, for both of single-layer BP and bulk BP, the calculated dielectric tensor elements show little difference between armchair and zigzag directions, illustrating the slight anisotropic screening property of BP.

Then for comparison, we calculate the dielectric responses, $\varepsilon(\mathbf{q}, \omega)$, both under the long-wavelength limit $q = 0$ Å$^{-1}$ and at $q = 1.1$ Å$^{-1}$. Fig. 2(a)-(b) show the imaginary and real parts of frequency-dependent dielectric function under long-wavelength limit. Static dielectric constant corresponds to the zero-frequency limit of the real part of the dielectric function [34]. At $\omega = 0$ eV, real parts of dielectric function under long-wavelength limit for two directions are slightly different, which is consist with the results in Table I. The imaginary part of the permittivity describes the dissipation of energy, with $-\text{Im}[1/\varepsilon(\omega)]$ proportional to the electron energy loss spectroscopy (EELS) loss function [35]. The peak for imaginary part of dielectric function is closely related to the dispersion of the plasmons versus momentum. As shown in Fig. 2(c)-(d), the frequency-dependent dielectric function curves at $q = 1.1$ Å$^{-1}$ show more obvious anisotropy than that at the long-wavelength limit, which is in agreement with previous results about the anisotropic dispersion of the plasmons versus momentum for BP [35]. The dielectric curves of single-layer BP in Fig. S2. also show similar behavior.

With a general understanding of the intrinsic slight anisotropic screening of BP, we then focus on the effects of doping on charge screening. Former studies show some evidences about the close connection of doping and charge screening. For some semiconductor materials, doping influences the dielectric functions and planar average of the charge density [25,26]. Position of dopant ion leads to the increase of screening efficiency for hydrogenic donor [36]. From Ref. [24], we can find a more direct relation about doping and screening of BP. Increasing K adatom density leads to the transition from isotropic to anisotropic screening on BP surface. All of these suggest that doping can affect the charge screening of BP. In Ref. [24], the authors considered the K doping range between $7 \times 10^{11}$ cm$^{-2}$ ~ $1.8 \times 10^{13}$ cm$^{-2}$ in experiments. In this work, we use doping degree range of $-9.6 \times 10^{13}$ cm$^{-2}$ ~ $3.2 \times 10^{13}$ cm$^{-2}$ for single-layer BP and $-4.8 \times 10^{20}$ cm$^{-3}$ ~ $3.6 \times 10^{20}$ cm$^{-3}$ for bulk BP. Negative doping concentration means electron doping and the positive one means hole doping. As shown in Fig. 3, doping degree changes would lead to obvious increase of $\varepsilon_a$ and $\varepsilon_b$. The dielectric tensor elements of BP are calculated from equation (3), and influenced greatly by the intrinsic band structures and occupations of each band in equation (1) [30]. The inter-band transitions mostly lead to the typical semiconductor-type charge screening of pristine BP. However, the occupations of bands near the Fermi level change with increasing doping and the intra-band transitions also start to play a role in the process. Therefore, BP shows metallic screening feature due to the enhancement of metallicity induced by charge doping.

It is worth noting that dielectric tensor elements of BP vary between anisotropy and isotropy with the increasing doping degree. In B. Kiraly *et al.*'s study, BP surface under low K adatom density ($n_k$) of $7 \times 10^{11}$ cm$^{-2}$ shows isotropy, and starts to change towards anisotropic behavior at medium density of $1.4 \times 10^{12}$ cm$^{-2}$ ~ $2 \times 10^{12}$ cm$^{-2}$, finally transforms to strong anisotropy at $n_k = 1.8 \times 10^{13}$ cm$^{-2}$ [24]. Similarly, our calculations about the single-layer BP indeed reproduce this trend. As shown in Fig. 3(a), dielectric tensor elements of single-layer BP exhibit slight anisotropy in the hole doping concentration range of 0 ~ $3.84 \times 10^{12}$ cm$^{-2}$. With further increase of hole doping degree, dielectric tensor elements then exhibit stronger anisotropy. In addition, the isotropic screening in Ref. [24] corresponds to the low degree of charge doping, as the experimental results were obtained with a certain degree of K adatoms. In Fig. 3(b), static dielectric tensors elements of bulk BP show anisotropy in 0 cm$^{-3}$ ~ $-1.2 \times 10^{20}$ cm$^{-3}$. As doping degree increases, dielectric tensor elements show nearly isotropy at the doping degree of $-1.8\times10^{20}$ cm$^{-3}$ ~ $-3\times10^{20}$ cm$^{-3}$ and get back to anisotropy later. Another interesting phenomenon is that relative dielectric tensor elements of two directions vary with doping concentration. As shown in Fig. 3(a), in the range of $-1.92\times10^{13}$ cm$^{-2}$ ~ $-3.2\times10^{13}$ cm$^{-2}$, relative value of dielectric tensor elements along zigzag direction is larger. When doping concentration increase to $-3.84\times10^{13}$ cm$^{-2}$, dielectric tensor element along armchair direction is larger than that along zigzag direction. Similar behavior also occurs in bulk BP in Fig. 3(b). As doping concentration increases, more reverses of relative dielectric tensor elements can be observed.

To further confirm the effects of doping on BP's charge screening, we study the charge density difference (CDD) near the single-layer BP surface induced by K adatom. The 2D charge redistribution patterns were plotted at different heights from the BP surface. Remarkably, the K adatom model used here already corresponds to an initial doping concentration, which is similar to the BP surface doped with charged K adatoms in Ref. [24]. In this case, initial doping concentration of $8.24\times10^{12}$ cm$^{-2}$ corresponds to the situation between medium ($2.0\times10^{12}$ cm$^{-2}$) and high degree ($1.8\times10^{13}$ cm$^{-2}$) of $n_k$ in B. Kiraly *et al.*'s work [24]. The CDD distribution with initial doping for single-layer BP shows anisotropy, which is in line with the experimental observations. Upper parts of Fig. 4 show the calculated CDD, and medium parts indicate the cross profile at the height of K adatom while the bottom parts correspond to the cross profile at the height of BP plane. One can see that electrons are transferred from the K adatom to the BP plane. With doping degree increases, the distribution range of CDD is narrowed, which is in agreement with the increased charge screening induced by doping shown in Fig. 3. Similar trends of CDD can be observed in the hole doping in Fig. S3. On the other hand, the shape of 2D section diagram also changes with doping level. Without doping, the 2D charge redistribution patterns exhibit obvious anisotropy, as shown in Fig. 4(a). With the increases of doping concentration, 2D section diagram first shows isotropic screening property, and then exhibits anisotropy when doping concentration increase to $-1.65\times10^{14}$ cm$^{-2}$. The calculated CDD further verify that the balance between isotropic and anisotropic screening could be tuned by doping. In Ref. [16], BP surface exhibits isotropic charge screening with the K adatom density of about $1\times10^{12}$ cm$^{-2}$. This K adatom density is located between the low density ($7\times10^{11}$ cm$^{-2}$) and medium one ($1.4\times10^{12}$ cm$^{-2}$) in Ref. [24], corresponding to the transition from isotropy to slight anisotropy. While the more obvious anisotropic charge screening was observed on BP surface with higher K adatom density in Ref. [24]. Therefore, the diverse conclusions in Ref. [16] and Ref. [24] about isotropic or anisotropic screening in BP may originate from the different K adatom densities. Our first-principles calculations shown here provide direct theoretical evidences for this influence of charge doping on the charge screening of BP.

## 4 Conclusions

In conclusion, on the basis of first-principles calculations, we study the screening properties in single-layer/bulk BP and the influences of charge doping on them. Without doping, the calculated static dielectric tensor elements and frequency-dependent dielectric functions of BP only show inconspicuous anisotropy. The electron and hole doping can significantly increase the screening in BP, resulting in a transition from semiconductor- to metal-type screening. With the change of doping degree, the relative strength between charge screening along zigzag and armchair directions changes too. Either isotropic or anisotropic charge screening can exist, under different levels of doping. We also investigate the 2D charge redistribution patterns induced by K adatom near the BP surface, which further confirms that the balance between isotropic and anisotropic screening could be tuned by doping. The unveiled slight anisotropic screening of BP and the significant role of charge doping in affecting screening will shed light on further study of the screening properties of BP. The sensibility of charge screening in BP to doping also make it a promising candidate for potential tunable and flexible narrow-gap compound semiconductors for infrared optoelectronics [37].

**Acknowledgements**

Yexin Feng is supported by the National Basic Research Programs of China with grant No. 2016YFA0300901 and the National Science Foundation of China with grant Nos. 11974105, 11604092, 11634001.

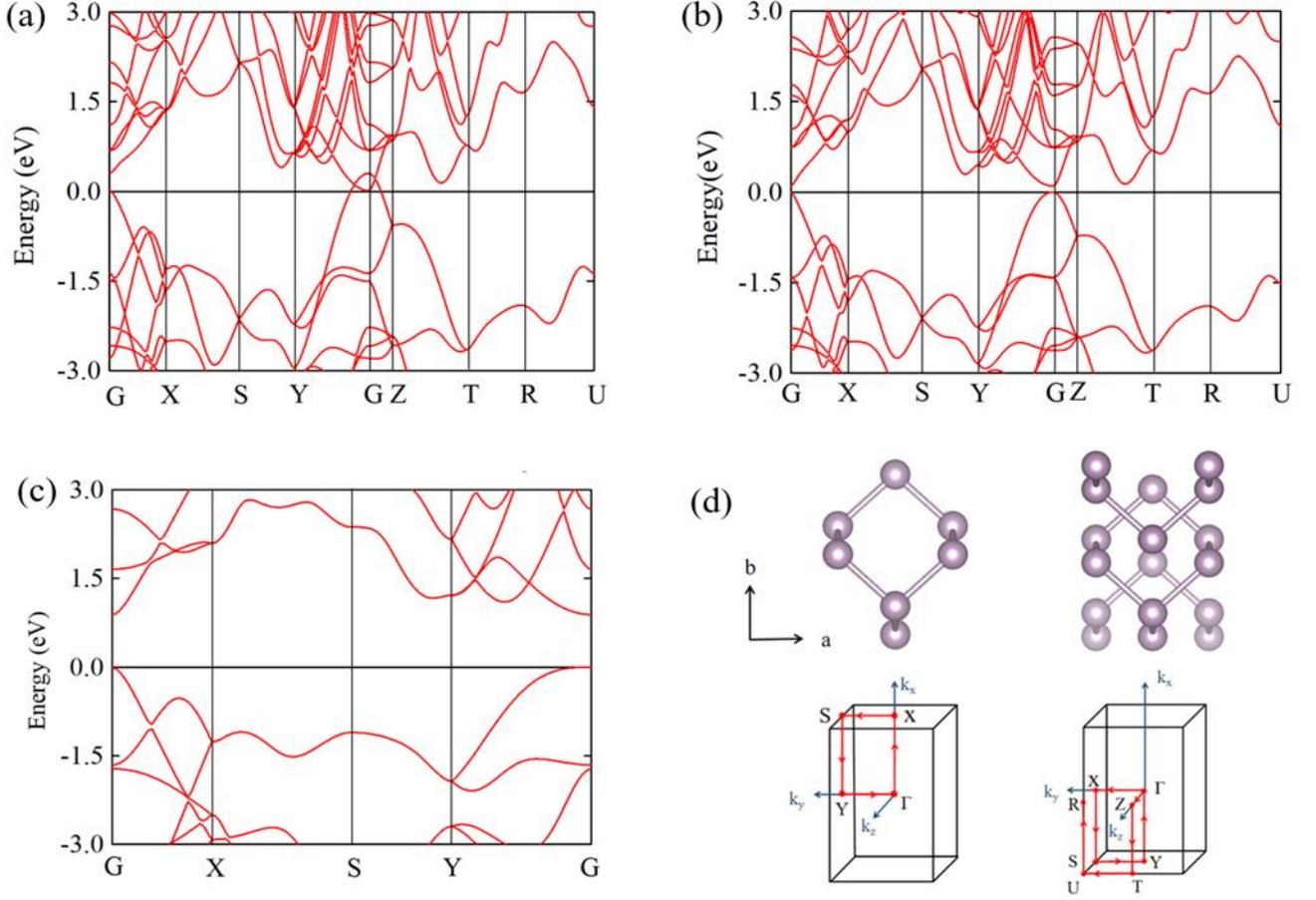

**FIG. 1.** The band structures of (a) bulk BP, (b) bulk BP (with 3% in-plane strain) and (c) single-layer BP. (d) Top view of crystal structures, Brillouin zones and the high-symmetry points of single-layer BP and bulk BP.

**Table 1.** The static dielectric tensor elements $\varepsilon_a$, $\varepsilon_b$ and $\varepsilon_c$ of single-layer BP, bulk BP and bulk BP with 3% in-plane strain.

| Systems | $\varepsilon_e^a$ | $\varepsilon_e^b$ | $\varepsilon_e^c$ |
|---|---|---|---|
| Single-layer | 1.03 | 1.06 | N/A |
| Bulk | 13.27 | 32.1 | 9.72 |
| Bulk (with 3% in-plane strain) | 14.96 | 17.05 | 10.15 |

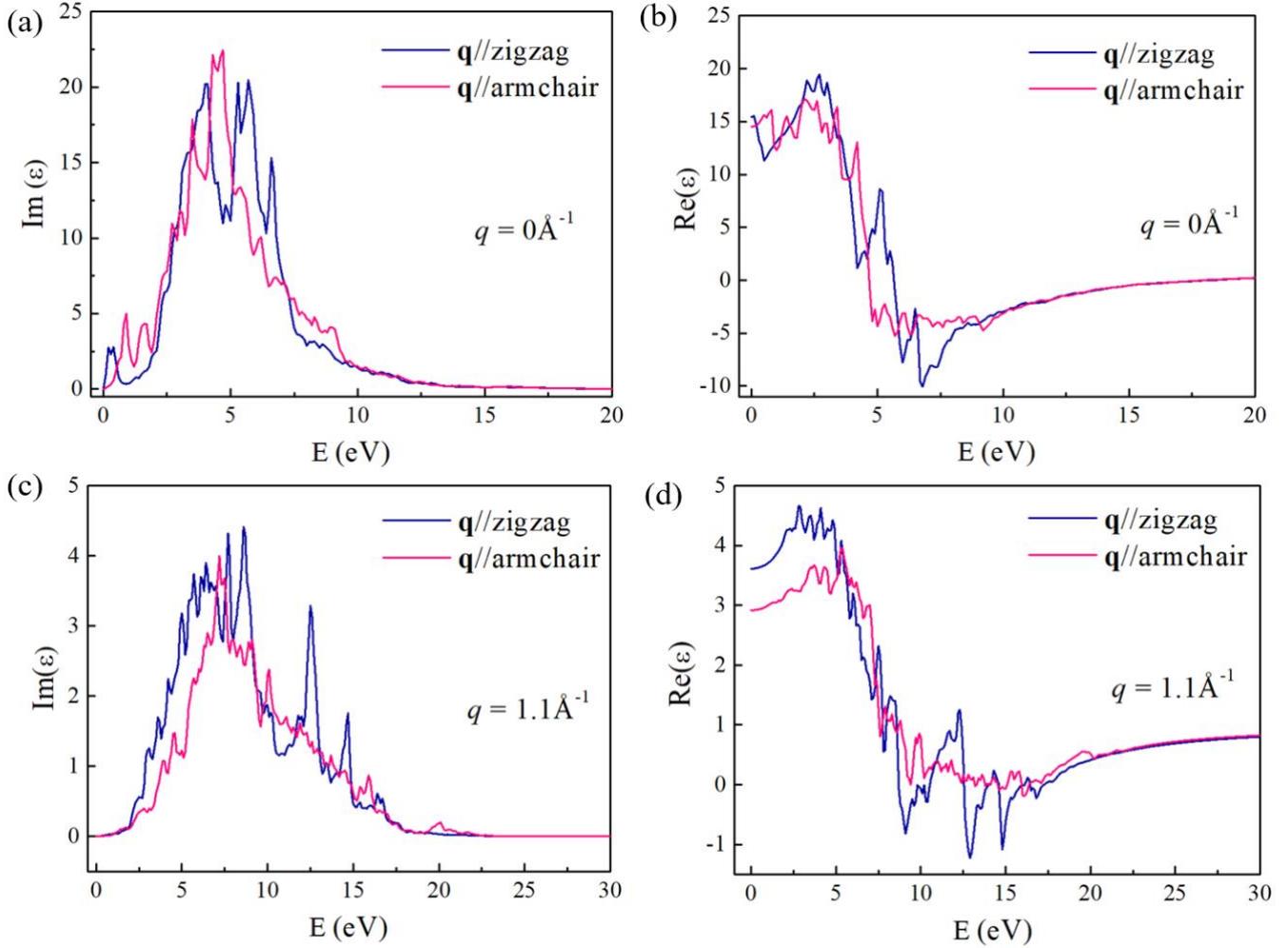

**FIG. 2.** (a) (b) The imaginary and real parts of frequency-dependent dielectric function at $q = 0$ Å$^{-1}$ for bulk BP (with 3% in-plane strain). (c) (d) The imaginary and real parts of frequency-dependent dielectric function at $q = 1.1$ Å$^{-1}$ for bulk BP (with 3% in-plane strain).

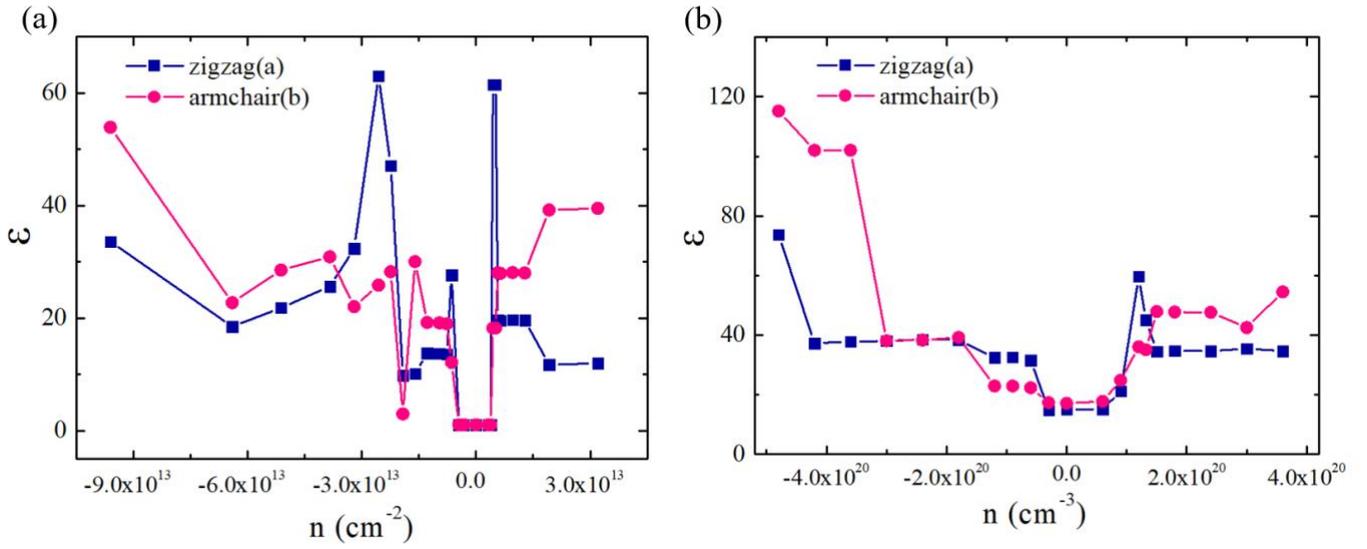

**FIG. 3.** (a) Calculated static dielectric tensor elements $\varepsilon_a$ and $\varepsilon_b$ for single-layer BP as a function of doping concentration. (b) Static dielectric tensor elements $\varepsilon_a$ and $\varepsilon_b$ for bulk BP (with 3% in-plane strain) as a function of doping concentration.

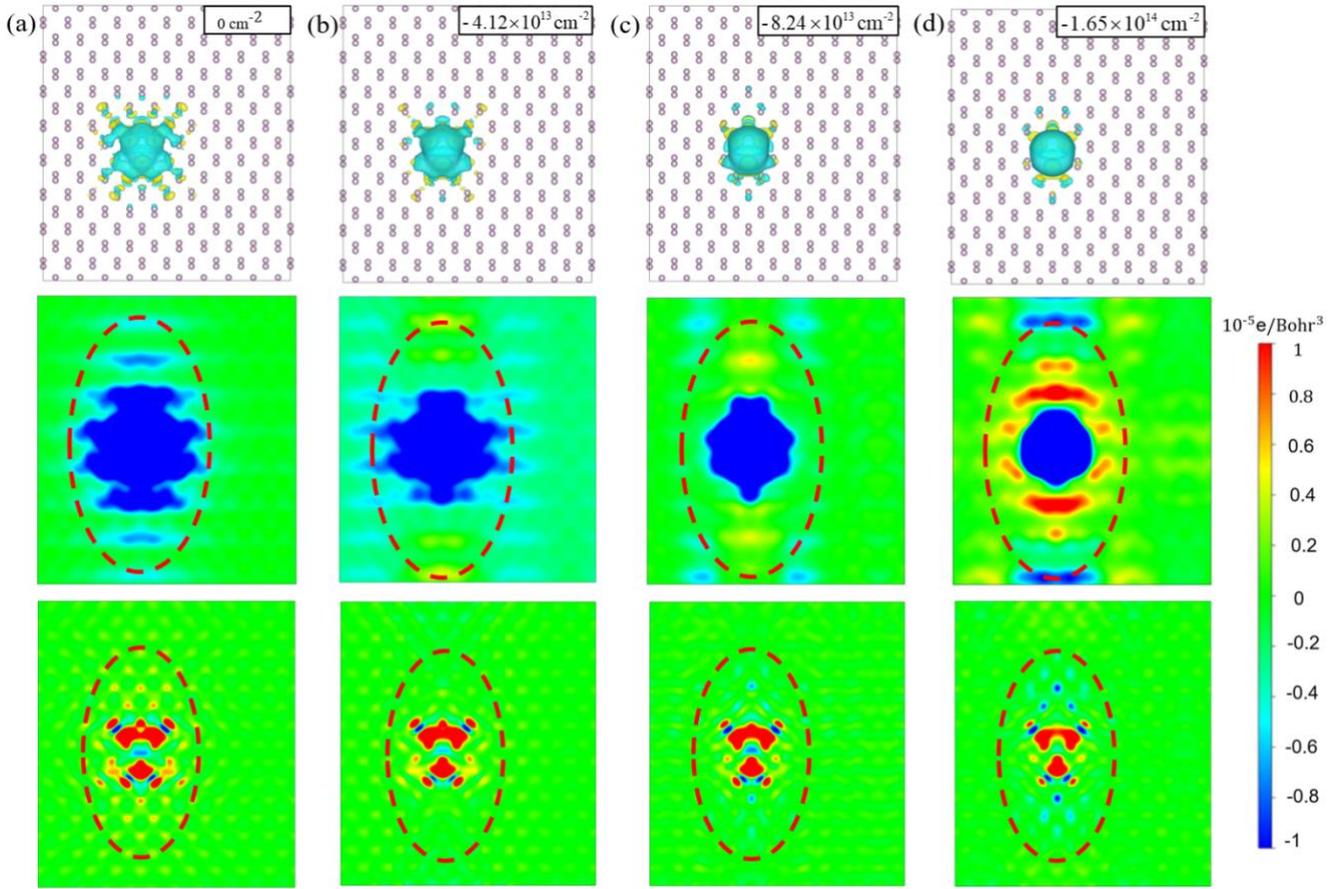

**FIG. 4.** The top view of charge density differences for single-layer BP with K adatom (top) and the 2D section diagrams of charge density difference at the height of K adatom (medium) and BP plane (down) under different electron doping concentrations: (a) 0 cm$^{-2}$ (b) –4.12×10$^{13}$ cm$^{-2}$ (c) –8.24×10$^{13}$ cm$^{-2}$ (d) –1.65×10$^{14}$ cm$^{-2}$. Positive and negative values of the charge density difference is shown by yellow and cyan parts respectively, with the isosurface value of ±0.00022 e/Bohr$^3$. The red dashed ovals are of the same size, which is a guide for comparison.

# Supplemental Material For

# Isotropic or anisotropic screening in black phosphorous: can doping tip the balance?

Zhimin Liu[1], Ye Yang[1], Yueshao Zheng[1], Qinjun Chen[1], Yexin Feng[1, †]

[1] Hunan Provincial Key Laboratory of Low-Dimensional Structural Physics & Devices, School of Physics and Electronics, Hunan University, Changsha 410082, People's Republic of China

Corresponding author. E-mail: [†]yexinfeng@pku.edu.cn

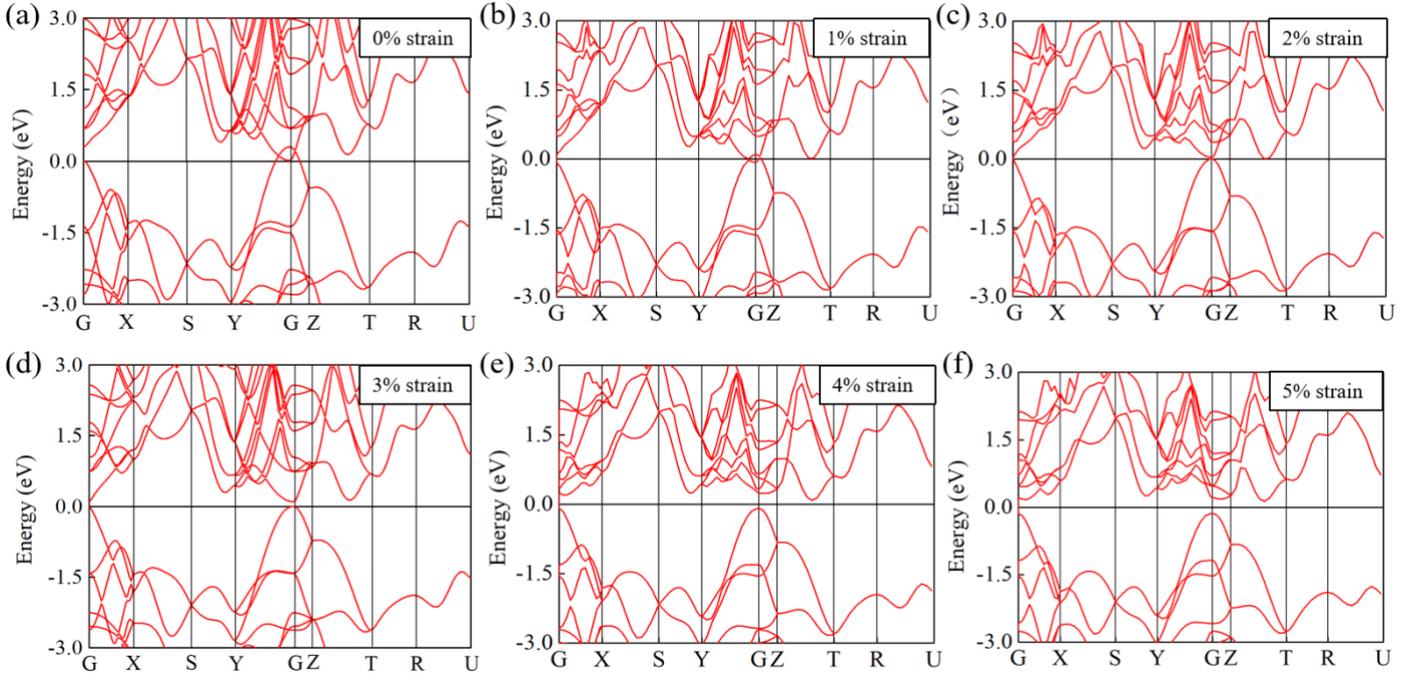

**FIG. S1.** Band structures of bulk BP with different in-plane strain.

| In-plane strain(%) | $\varepsilon_a$ | $\varepsilon_b$ | $\varepsilon_c$ |
|---|---|---|---|
| 0 | 15.9 | 655.94 | 11.54 |
| 1 | 15.51 | 77.19 | 11.05 |
| 2 | 15.13 | 1037.95 | 10.70 |
| 3 | 14.96 | 17.05 | 10.15 |
| 4 | 14.74 | 16.58 | 10.09 |
| 5 | 14.67 | 16.16 | 10.07 |

**Table S1.** Static dielectric tensor elements along *a*, *b* and *c* direction under different in-plane strain with experimental lattice constants.

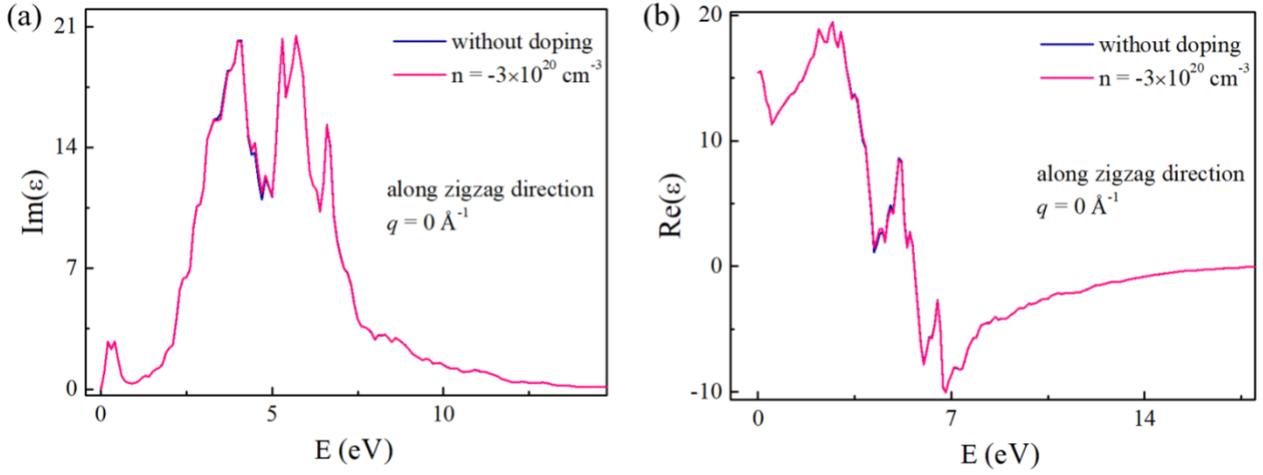

**FIG. S2.** Imaginary (a) and real(b) parts of frequency-dependent dielectric functions for bulk BP without doping and under doping concentration of $-3\times10^{20}$cm$^{-3}$.

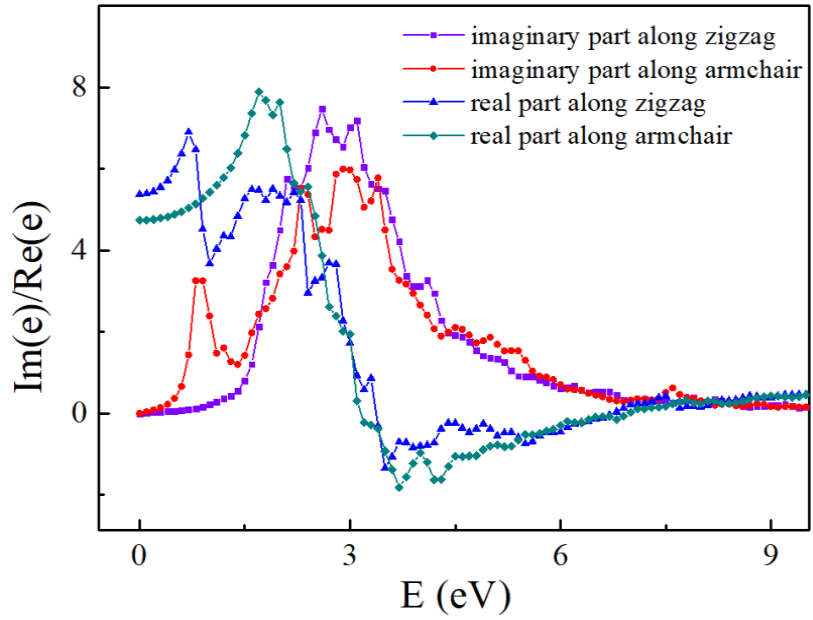

**FIG. S3.** Imaginary and real parts of frequency-dependent dielectric function for single-layer BP along zigzag and armchair directions under long-wavelength limit.

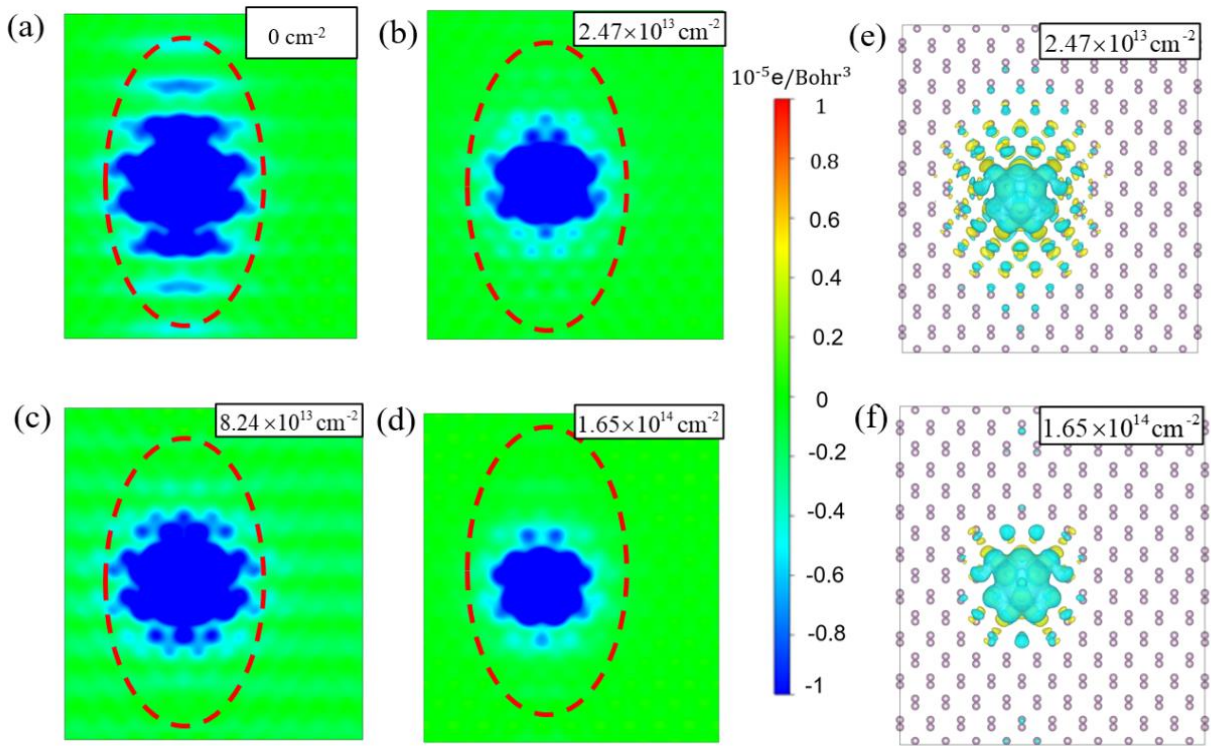

**FIG. S4.** 2D section diagrams of charge density difference for single-layer BP with adsorbed K at the height of K adatom, under different levels of hole doping: (a) 0cm$^{-2}$, (b) 2.47×10$^{13}$cm$^{-2}$, (c) 8.24×10$^{13}$cm$^{-2}$, (d)1.65×10$^{14}$cm$^{-2}$. The top view of charge density difference (with the isosurface value of ±0.00022 e/Bohr$^3$) for single-layer BP with K adatom under different electron doping concentrations: (e) 2.47×10$^{13}$cm$^{-2}$ (f) 1.65×10$^{14}$cm$^{-2}$.